\documentclass[runningheads]{llncs}

\usepackage[T1]{fontenc}
\usepackage{graphicx}
\usepackage{tabularx}
\usepackage{rotating}
\usepackage{multirow}
\usepackage{xcolor} 

\newcommand\infocategory[1]{\textcolor{teal}{#1}}
\newcommand\action[1]{ $\diamond$~\textcolor{magenta}{#1}}

\begin{document}
\title{Evaluating Actionability in Explainable AI}
%
%
\author{Gennie Mansi\inst{1}\orcidID{0000-0001-6186-5102} \and
Julia Kim\inst{1}\orcidID{0000-0002-8644-9321} \and
Mark Riedl \inst{1}\orcidID{0000-0001-5283-6588}}

\authorrunning{G. Mansi et al.}
%
\institute{Georgia Institute of Technology, Atlanta GA 30332, USA \\
\email{gennie.mansi@gatech.edu, julia.kim@gatech.edu, riedl@cc.gatech.edu}}
\maketitle              
\begin{abstract}
A core assumption of Explainable AI (XAI) is that explanations are useful to users---that is, users will \textit{do} something with the explanations. Prior work, however, does not clearly connect the information provided in explanations to user actions to evaluate effectiveness. In this paper, we articulate this connection. 
We conducted a formative study through 14 interviews with end users in education and medicine. We contribute a catalog of information and associated actions. Our catalog maps 12 categories of information that participants described relying on to take 60 different actions. We show how AI Creators can use the catalog's specificity and breadth to articulate how they expect information in their explanations to lead to user actions and test their assumptions. 
We use an exemplar XAI system to illustrate this approach. We conclude by discussing how our catalog expands the design space for XAI systems to support actionability.

\end{abstract}
\section{Introduction}
Artificial intelligence (AI) systems are increasingly used by people in high-stakes decision making for domains such as healthcare, housing, financial, and educational determinations~\cite{Lipton2018,ChariEtAl2020,GuidottiEtAl2018,DohnEtAl2020}.
To support these people, explainable AI (XAI) systems relay explanations about the AI system to help users understand and act on the AI systems' outputs ~\cite{Lipton2018,ChariEtAl2020,GuidottiEtAl2018}. 
For example, a doctor may use an XAI system to understand and confirm if a medical AI system's recommendation aligns with medical standards, so they can prevent unexpected harms. 

Evaluations play a critical role in helping AI creators know whether explanations are effective.
Some XAI evaluations assess how well a XAI algorithm explains an underlying AI model. For example, algorithmically-centered evaluations can ensure an XAI method accurately reflects the major features of an AI model that impacted a decision; they can also reflect the consistency an XAI algorithm's performance given different inputs \cite{CoroamaAndGroza2022}. 
Qualitative evaluations are used to verify and compare how non-AI expert users of AI systems perceive the readability, intuitiveness, and clarity of different methods \cite{CoroamaAndGroza2022}.

As the concept of explainability has progressed, researchers highlight that AI systems must go beyond communicating why or how a decision was made to improve transparency and decision making.
Explanations must be \textit{actionable}---enabling people to take pragmatic actions in response to AI systems and decisions
\cite{JornoAndGynther2018,LyuEtAl2016,ChoEtAl2019,JoshiEtAl2019,TanAndChan2016,WiratungaEtAl2021,SinghEtAl2021}.
However, a growing number of studies show that some AI explanations are not actionable for end users, such as doctors and teachers, using AI systems in their workflows \cite{LiaoEtAl2020,CortiEtAl2024}. 
This suggests a need to improve the XAI community's ability to articulate and evaluate how information in their explanations supports actionability for users in real-world settings.

However, no consensus yet exists on the information that can---or should---be presented to users to effectively enable user actions. 
While several papers~\cite{Lipton2018,GuidottiEtAl2018,Doshi-VelezAndKim2017,WangEtAl2019,AryaEtAl2019,ICOAndTuring2022,ChariEtAl2020} attempt to categorize information that explanations communicate, they do not elaborate expected user actions. 
Others evaluate explanations on users' ability to act on unexpected behaviors \cite{ChoEtAl2019}, change an outcome \cite{MothilalEtAl2020,JoshiEtAl2019,ChiangAndDey2018}, or achieve a desired result \cite{SinghEtAl2021,EnsanEtAl2017,GhoshEtAl2018} in the context of specific technical solutions.
To our knowledge, user actions taken in response to explanations have not been holistically considered in a single analysis. 

Clearly connecting how people use information in explanations to take actions will help AI Creators evaluate and improve the actionability of their systems. This paper improves XAI evaluations by answering the following questions:
\begin{enumerate}
    \item \textbf{RQ1} - What actions do users take to leverage explanations in their decision making?
    \item \textbf{RQ2} - What kinds of information do users seek from explanations to support their actions? 
   \item \textbf{RQ3} - How do evaluations of actionability highlight design opportunities to better support users?
\end{enumerate}
We answered these questions by conducting a formative study with 14 participants in two domains, medicine and education, chosen because of their current use of XAI systems in high-stakes, multi-stakeholder decision-making environments. 
We identified and mapped 60 user-defined actions that relied on 12 categories of information, creating a catalog that can be used to evaluate the actionability of XAI systems. 

The specificity and breadth of the actions and information in our tool is a novel contribution that guides evaluation in two ways.
First, it helps AI Creators articulate how they expect information in their explanations to lead to specific user actions. 
Second, by articulating their expectations, AI Creators can then conduct measurable, user-centered evaluations to test their assumptions. 
To understand how their systems can better align with user needs, AI Creators can look for how users {\em are} or {\em are not} acting as intended. 
In the rest of the paper, we detail how we created the tool, and we present the kinds of information and actions in the tool. We provide an extended example of how system developers can apply our catalog to create evaluations for an XAI system with a ``black box'' model, and then guide development accordingly. Finally, we discuss how our tool provides insights into
design opportunities for XAI, and more broadly AI, communities to better evaluate with and support users around actionability.

\section{Related Work} \label{RelatedWork}
AI systems are being created for and deployed in critical areas, including healthcare, education, housing, and finance \cite{Lipton2018,ChariEtAl2020,GuidottiEtAl2018,DohnEtAl2020}. To support humans’ decision making, trust, and agency in these critical areas, XAI systems provide human-understandable explanations for an AI's reasoning or response \cite{JornoAndGynther2018,ChoEtAl2019}. AI Creators have explored a variety of methods to relay information about the system in a way that users can understand.

One approach is to create inherently explainable AI systems, built on underlying algorithms that are easy for humans to understand. These systems provide information about how they are built, so people can check system decisions and act to enable positive outcomes for themselves \cite{GuidottiEtAl2018}. Increasingly, there are ``Black Box'' AI systems with an opaque internal state that work best in application but do not have inherently explainable algorithms that are easy to understand. Consequently, post-hoc techniques have been developed to help people understand less transparent systems' behaviors while remaining opaque about the algorithm's implementation \cite{GuidottiEtAl2018,GunningAndAha2019}. Together, inherently explainable and post-hoc methods produce a variety of explanations that can be categorized broadly into those that provide explanations given specific inputs (e.g. the training dataset used) versus those providing system-level information (e.g. accuracy and coherence measures) \cite{EdwardsAndVeale2017,GuidottiEtAl2018}. 

Evaluations are crucial for helping AI creators understand their implementation of XAI systems because they reflect explanations' effectiveness and identify areas for improvement. For inherently explainable and post-hoc methods, technical features, such as faithfulness, are used to perform benchmarks assessing relative performance \cite{AdadiAndBerrada2018,ArrietaEtAl2019,MohseniEtAl2020}. For example, faithfulness reflects to what extent the explanation corresponds to the actual variables impacting the model's output, ensuring important model features are actually captured and reflected \cite{CoroamaAndGroza2022}. 
Qualitative assessments are also often used to verify and compare the readability, intuitiveness, and clarity of different algorithmic advancements \cite{CoroamaAndGroza2022}, calling attention to how non-algorithmic users of AI systems may perceive an explanation.  

As the concept of explainability has evolved, researchers have emphasized that to achieve transparency and accountability, AI systems do not just need to communicate why or how a decision was made, they also need to enable pragmatic action~\cite{JornoAndGynther2018,LyuEtAl2016,ChoEtAl2019,JoshiEtAl2019,TanAndChan2016,WiratungaEtAl2021,SinghEtAl2021}. That is: they need to be {\em actionable}. \textbf{Actionability} refers to the ability to support pragmatic action and is closely tied to aspirations around human oversight and improved decision making, trust, and agency \cite{AndradaEtAl2022,GliksonAndWoolley2020}. 
Studies demonstrate that despite performance in algorithmic evaluations, non-technical users report limited actionability for XAI systems deployed in real-world settings \cite{LiaoEtAl2020,CortiEtAl2024}, indicating a disconnect between algorithmic evaluation methods and the ability for explanations to support actionability.

Evaluations that connect information in explanations to actions that users take could help AI Creators close the gap around actionability in real world settings. 
In computer programming contexts, Cho et al.~\cite{ChoEtAl2019} analyzes actionability in terms of whether explanations help programmers fix unexpected program behaviors. In healthcare, the actionability of explanations has been cast in terms of revealing actions clinicians can take in AI-based health record systems \cite{OngEtAl2021,BharambeAndSrinivasaraghavan2021}, allowing radiologists to find anatomical locations from an AI-generated report \cite{RobertsEtAl2012}, and highlighting behavior changes in personalized health feedback for patients \cite{ChiangAndDey2018}. In education, actionability has been used to refer to AI-based learning analytics tools that allow learners and other stakeholders to take actions that improve classroom outcomes \cite{RoseEtAl2019,JornoAndGynther2018}, including self-reflection \cite{OuyangAndJiao2021}, trust \cite{SusnjakEtAl2022,KhosraviEtAl2022}, and agency \cite{OuyangAndJiao2021,KhosraviEtAl2022}. Still others, agnostic of domain, define actionability in terms of the ways they allow users to change an outcome \cite{MothilalEtAl2020,JoshiEtAl2019} or achieve a desired result \cite{SinghEtAl2021,EnsanEtAl2017,GhoshEtAl2018}. 

Evaluations need to adapt to our evolving understanding of explainability. As indicated by the diversity of actionability evaluations above, there is no systematic catalog of actions and information to help AI Creators compare different kinds of explanations with users. 
This limits our understanding of the actions that users desire and the algorithmic developments to support these users.
Evaluating XAI systems based on users' actions directly and clearly reflects XAI's core values around actionability.
In our study, we synthesize a catalog of actions and information---created with end users---that serve as a resource for AI Creators to draw on when evaluating their systems.

\section{Methods}
To help AI Creators evaluate the actionability of their explanations, we used qualitative empirical and design methods with doctors and teachers to systematically catalog their actions and information needs. We conducted interviews within two professions, education and medicine, because they involve high-stakes decisions that meaningfully effect people's lives and society as a whole; commercial AI developers are actively developing AI tools for these fields; and conducting research with two professions enabled us to look for commonalities and the potential transferability of our findings.

We used a scenario-based design activity in our interviews. Scenario-based designs (SBD)\cite{Carroll1995} are used to understand non-technical users' needs from AI-based decision support systems (e.g. \cite{EE_EhsanEtAl2021,Seam_EhsanEtAl2024,WolfAndBlomberg2019}).
Non-technical users often cannot describe the exact algorithms or technical specifications they need, but they can articulate what actions they want to take with a technical system during a familiar context or task. For example, a doctor may articulate wanting to find out if and how drug contraindications were considered in an AI determination, without using the corresponding technical terms in algorithm development such as ``variables'' or ``parameters'' that they do not know. 
SBD prompts users to ground their approach in their work environments and how different aspects of that situation impact their ability to \textit{act} effectively.
Because SBD focuses on getting participants to describe their actions, rather than technology requirements or solutions, SBD does not require participants to have technical knowledge of AI \cite{Ackerman2000}.

Per SBD’s method~\cite{Carroll1995}, our interview protocol immersed participants in a profession-specific scenario, using a visual to ease their cognitive load. Each scenario was designed to provide contextual specificity that could be interpreted based on individual differences, to include experience levels and socio-economic status. 
For the doctors, the scenario put them in the role of the attending physician at a large hospital who uses an AI-based recommendation system to decide on drug dosages for a patient who is also receiving care from other doctors. To match their context, doctors' visual was designed as a mock electronic-health record (EHR) interface.
For teachers, our scenario put them in the role of using an AI-based recommendation system to place students into different levels of computer science courses with the help of other school staff. 
Both scenarios involved decisions impacting people who are not direct users of the AI system: patients and students, respectively. 

The visuals were designed to elicit actions without confusing or overwhelming participants with technical jargon. As per best practices, the visuals incorporated familiar, realistic buttons and features to help generate ideas based on the real-world situation; they reflected potential actions and information relevant for professionals in each domain, such as requesting an audit or referencing a patient or student information.
The interface was intentionally non-interactive---additional screens or windows did not appear.
Instead, we asked participants to describe the expected or desired behavior of an interface element. When necessary, we prompted participants to state the helpful and unhelpful interface elements for achieving their goals in the scenario; the parts they would change and how; and what kinds of actions they could take based on the information in each component. 

We recruited physicians and teachers for two rounds of pilot studies to finalize the scenarios, interview procedure, and designs of the visuals. The pilots confirmed that the use-case scenarios and visuals corresponded to real world environments. 
We refined the visuals based on feedback, reducing the amount of information and adding more contextually-relevant resources. 

\subsection{Study Procedure}
We obtained IRB approval for the research and informed consent from all participants. Interviews were conducted online with screen-sharing for the interviewees and lasted 72 minutes on average. Our interview procedure was as follows. 

First, we asked participants to imagine themselves in the scenario and to describe the kinds of tasks they might need to accomplish, people they might need to talk to, and goals they might have. Then we presented the visual, and participants were then asked to think out loud about the ways they would approach their decision-making with it. This included whether and how they would incorporate and act on the visual’s elements and other real-world information. Participants were instructed to rely on their past and present experiences to explain the types of information they would use, the way they would use them, and the reasons for those choices. Participants explored and engaged the visual with minimal prompting.

\subsection{Recruitment} \label{recruitment}

Recruitment was initiated through online messaging and contacts of the research team, followed by snowball sampling. For doctors, the recruiting criteria were: must have an MD and at least be in their medical residency. For teachers, the recruiting criteria were: must teach computer science at the middle school or high school level and have at least 1 year of teaching experience. We verified these criteria were met through correspondence with participants prior to interview. 

We recruited a total of 9 doctors (labeled M1-M9) and 5 teachers (labeled T1-T5) with a range of time-in-service and expertise. 
Doctors ranged in their expertise from first year residents to clinicians with 30+ years of experience. M6 and M9 had extensive experience teaching junior doctors and peers. M4 had a PhD in research with a clinical focus.  
Teachers ranged from 7-30 years of teaching experience.  One teacher taught the AI pathway for a local high school. Two teachers taught AI units as a part of their classes. The other two teachers did not teach AI as a part of their curriculum but were interested in doing so.

\subsection{Analysis} \label{analysis}
All interviews were video-recorded to include the interface activities, then transcribed and anonymized for analysis. The first and second authors conducted an iterative thematic qualitative analysis of our data. They identified the information that participants wanted and the actions taken with that information. As we conducted our qualitative analysis, we found users’ terminology to describe AI systems diverged from those traditionally used in AI literature, leading to ambiguities in the codes. To address this, we created a list of key terms referred to by users and then re-analyzed the quotes that used the terms to derive user-centered definitions. We then revisited the original quotes, iterating on definitions until both coders agreed they reflected participants' meanings. Section  \ref{Appendix_Terminology} details the final set of user-based terms and definitions used to code the data. 
To support our research goals and findings, we define and use the user-derived terminology within the remainder of the paper. 

For the information types that participants wanted, the first and second authors produced in-vivo codes, which they thematically grouped to produce higher-level codes.
We then re-coded a sub-selection of the quotes using the higher-level codes. For codes that could not be clearly applied, we revised the definitions then re-coded a different sub-selection of the quotes. We iterated on the information codes until we reached a consensus and could re-code all selected quotes without conflict. This resulted in a set of \textit{Information Category} codes that describe the information that participants discussed. The information categories were then classified into 4 thematic groups, reflecting related kinds of information that users seek from the AI system to take actions.

We followed the same process to derive \textit{Actions} that participants took with the information. We coded actions that participants explicitly stated they would take or want to take when using the interface to make a decision. We categorized the actions into 3 groups, reflecting different dimensions of users' actions, which are detailed in Section \ref{ActionTypes}.

After defining the user-generated information categories, actions, and thematic groups, the first author re-coded the entire dataset. We selected approximately 20\% of the quotes in each interview for the second author to re-code for inter-rater reliability (IRR). This resulted in the selection of 160 of the 737 total quotes, or just under 22\% of total quotes. The inter-rater reliability for our study was 83.5\% according to Cohen's Kappa. In the case of disagreement, both authors discussed the application of the codes to selected participants' quotes until consensus was reached. After reaching consensus on the selected quotes, the entire dataset was re-coded again. 

\section{Findings \& Results} \label{Findings_Results}
Through our interviews, we created a catalog of user-generated actions and information to help AI Creators evaluate the actionability of their explanations. Our catalog is built on \textbf{User-Centered Terminology}, terms and definitions derived from our interviews. 
As noted in Section \ref{analysis}, participants' terminology around AI systems diverged from those used by the research community. 
AI Creators can use these terms from our catalog to aid clear communication with users.
We present here the participant-derived terms and definitions to facilitate discussion of our findings (full details in Appendix Section~\ref{Appendix_Terminology}).

Participants used {\bf\em AI system} (or \textit{system} for short) to refer to the AI algorithm \textit{and its corresponding user interface}. This includes explanations and also elements such as a ``contact us'' button that are not directly tied to the AI algorithm. 
Participants referred to the {\bf\em AI Creators} as all people who were involved in making the AI system's model \textit{and interface}. This includes software developers, the design team, and any other consultant who helped create the system, not just the AI model creators.
Because participants do not differentiate between the AI algorithm, explanations, and other interface components, we refer to any user interaction with the system as an {\bf\em AI Interaction}---rather than separating them into more specific categories like XAI, AI, or System Interactions.
In Section~\ref{Extended_Example}, we discuss how the AI community can apply these user-centered concept to evaluation and improve their systems. 

Participants referred to the person directly interacting with the AI system and making {\bf\em decisions}, or choices, using the AI system as the {\bf\em user}. Participants referred often to all people impacted by the user's decision-making and actions in response to AI system's outputs, which we term {\bf\em stakeholders}. Participants referred to the {\bf\em outcome} as the situations or consequences that result from users and/or stakeholders taking action with an {\bf\em AI output}. 
For example, the AI's recommendation of a 30 mg Penicillin dose is the \textit{output}; the doctor who directly uses the AI system is the \textit{user}, and their choice to prescribe a 50 mg Penicillin dose is the \textit{decision}; the patient and their caregivers are both \textit{stakeholders}, and the patient's reduced pain is the \textit{outcome}.

Next, we present the two building blocks for conducting evaluations---\textbf{User-Centered Actions} and \textbf{User-Centered Information Categories}---which we uncovered through the interviews and analysis.\textbf{ AI Creators can reference the actions and information together to articulate how each kind of information in their explanation should lead to different user actions.} This focuses evaluations on the actionability of the information in explanations.

\subsection{User-Centered Action Types}\label{ActionTypes}
One pillar of our catalog is User-Centered Action Types. 
From the interviews, we created a set of 60 User-Centered Action Types, which we organize into three broad dimensions of users' actions---AI Interactions, External Actions, and Mental State Actions. 
These dimensions can be used to understand the variety and breadth of actions users may want to take in response to the AI system. 
We introduce the three dimensions here with examples.
Appendix Section~\ref{Appendix_ActionCodes} has the full list of actions and their definitions.

\subsubsection{AI Interactions} Participants mentioned direct interactions with the AI system, such as clicking open a patient record (M1-M9) or student file (T1-T5), requesting information about the AI system (T1-T5, M1-M9), and messaging other doctors (M2-M5, M7, M9). These actions are categorized as \textbf{AI Interactions}.

\subsubsection{External Actions} Participants mentioned actions in the physical world such as speaking to another teacher (T1, T4), calling the on-call doctor or pharmacist (M3, M5, M7), or speaking to a caregiver about a decision (T2, T4). These actions---in response to, and outside of, the AI system---are categorized as \textbf{External Actions}.

\subsubsection{Mental State Actions} Finally, participants mentioned making choices or changing their beliefs as a result of the information provided by the system. 
For example, M1 said the following when asked what actions they would take from the information they wanted to see about the variables the model used, ``I'm trying to think of actions... I think I would trust the application more. I know that's not an active verb. But I would feel better [about the output].'' M1's actions are not physical interactions with the world nor are they interactions with the AI. Instead, their action is a \textit{change in belief} about how trustworthy the model is. 
Similarly, T3 said information about the model would help them have ``a better understanding of what a machine would be looking for'' to make a decision, and that it could help someone ``persevere and set a different goal.'' 
Along the same lines, M6 described how they would use information about the variables in the model to brainstorm alternative treatments for patients, ``Particularly if this patient has tried Gabapentin in the past and it's had a side effect or not been tolerated for some reason. So I would think that the AI might prioritize presenting multiple possibilities where they're available.'' M4 and M8 both described getting frustrated based on incorrect justifications or information in the system and deciding not to use the system in response. 
Other participants' actions included setting expectations about the model outcome (M1, M2, M5, M7, M9, T1, T4), understanding the AI Interactions and External Actions they could take (M1-M9, T1-T4), understanding the model (M1-M9, T1-T5), and validating information presented in the model (M1-M9, T1-T5).

Though they are not physically observable, brainstorming different options, understanding what the model is looking for, persevering, getting frustrated, and setting different goals are direct results of the explanation that \textit{users view as a valuable actions} to pursue a desired outcome from working with the AI system. 
Further, across all participants, Mental State Actions constituted the mode, with a total 2190 Mental State Actions taken in response to explanations, almost 10 times greater than the 200 AI Interactions and 270 External Actions they would take. 
Consequently, while AI Interactions and External Actions are the most easily observed and are often included in XAI's definitions of actionability \cite{JornoAndGynther2018},
we include \textbf{Mental State Actions} as a third, critical dimension of actionability. 

\subsection{User-Centered Information Categories and Associated Actions} 
\begin{figure}[t]
  \includegraphics[width=1\textwidth]{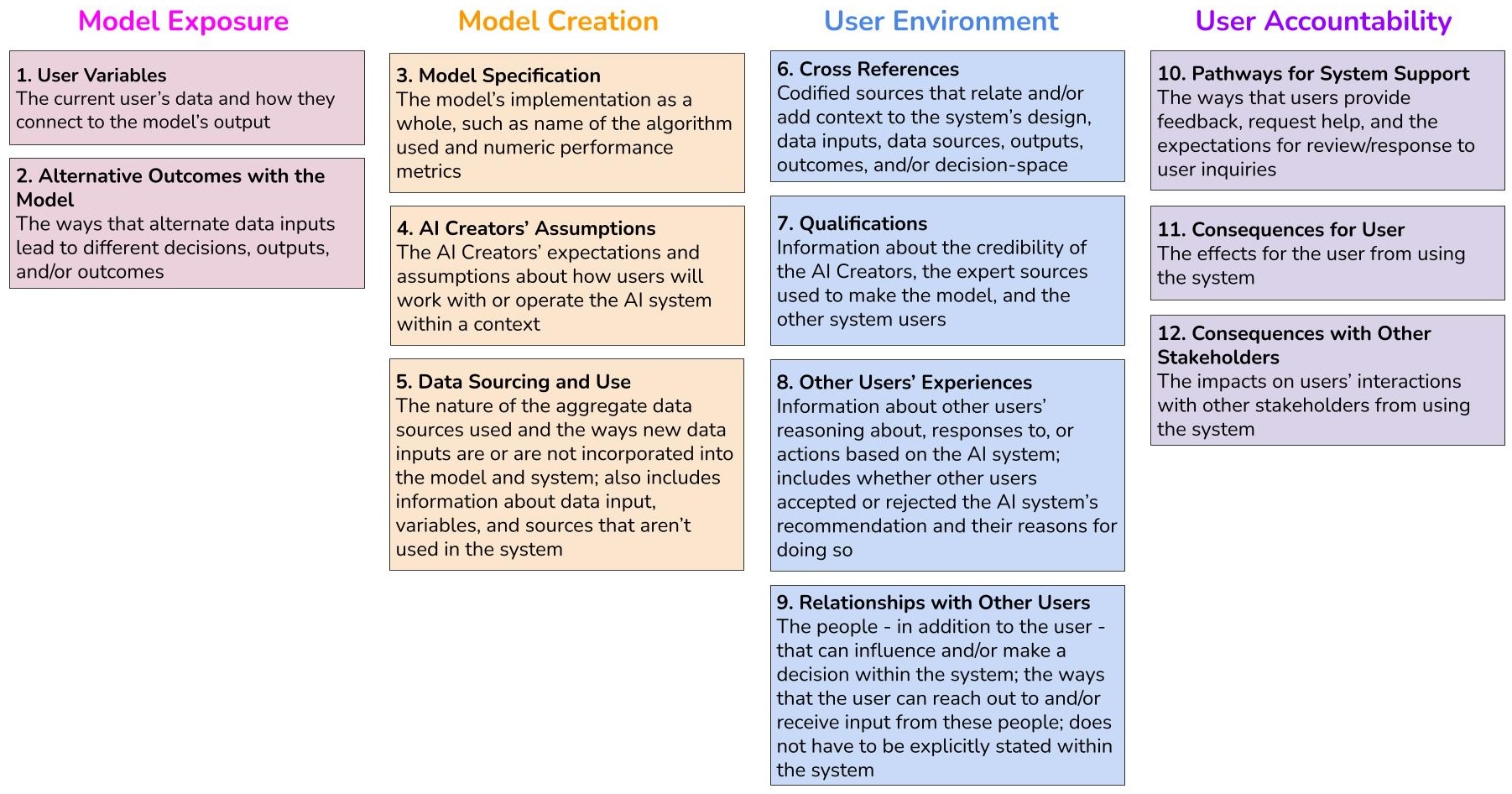} 
  \caption{Information Categories and their definitions categorized by theme.}
 \label{fig:CodingFrameworkOverview}
\end{figure}

The other pillar of our catalog is a set of User-Centered Information Categories.
The Information Categories reflect the range of the AI system's information that users draw on to take action. 
We present here the 12 participant-derived Information Categories.
For each Information Category, we show a selection of actions from the catalog that users took once given that information. 
Taken together, AI Creators can articulate how information in their explanations should lead to user actions. 
For readability, we use teal to denote \infocategory{Information Categories} and a different font to denote \action{user actions}. 
All 12 categories are shown in
Figure \ref{fig:CodingFrameworkOverview}, grouped by four themes. 
All Information Categories and associated Actions are included in Appendix~\ref{Appendix_FullFramework}.

\subsubsection{Information Categories related to Model Exposure} \label{Theme_ModelExposure}
The two Information Categories under the theme of Model Exposure reflect participants' desires to understand how system data is connected to the model's decision or different outcomes. 

\paragraph{\infocategory{\textbf{User Variables}}} Participants sought information from the AI system to connect \textit{the model's variables} with \textit{the model's output}. 
For example, M1 wanted the system to provide information that ``this person's glucose is out of range - here are drugs that, based on their age, would be applicable.'' In this example, two variables, glucose levels and age, are connected to the decision recommending drugs. 
M2 notes information from this category in the user interface, explaining ``[The interface] tells me what variable was analyzed, the medications and things like that,'' to contextualize the AI's decisions.

Mental State Actions were most commonly taken with this Information Category. 
Participants acted on information about the model's variables to \action{Recognize potential choices, options, and actions}, \action{Compare outputs from the current model against other knowledge}, and \action{Understand broader information about factors impacting the current environment or explanation}, among other actions. 
The top External Actions for this Information Category involve interacting with another person: 
\action{Communicate with another person}, \action{Consult another person about a decision}, and \action{Support another person's actions}. 
This the first of a recurring finding that Information Categories are used to engage and act with other people.

\paragraph{\infocategory{\textbf{Alternative Outcomes with the Model}}} Users sought information about how alternate data inputs (e.g. changes in students' test scores) lead to different decisions, outputs, or outcomes by or from the AI System. In our interviews, participants wanted this information to \action{Know possible outcomes} (M1, M2, M6, M7-M9, T1-T4) and \action{Change a decision about an action} or \action{Request system support} (M1, M3, M7-M9, T1, T4). 
This is the first instance of another recurring finding across our study: users want information from the system to \textit{predict or change the outcomes that could occur}.

\subsubsection{Information Categories related to Model Creation} \label{Theme_ModelCreation}

Three Information Categories under the theme of Model Creation provide the user with context about how the model was made and the algorithmic functioning of the AI, such as information about data inputs, assumptions, and specifications.

\paragraph{\infocategory{\textbf{Model Specification}}}
Participants sought technical specifications about the model, including the name of the AI algorithm used, descriptions of how the model works, and quantitative performance metrics. 
Specific to performance metrics, T5 expressed that just knowing the model was accurate and reliable based on the value of the accuracy ``will motivate you to use the use the information.'' 
Conversely, T1 didn't understand what accuracy meant or what an acceptable value for accuracy might be, but they still translated the value in terms of its impact on them if they did not receive a favorable outcome. They wondered, ``So I guess 95\% [accuracy] is good. I would hope it would be 100\%, right? But what happens to the other 5\% that are not correct? How many of them are actually able to do any of the steps [to improve] or have the information or know how to correct it?'' 
This quote demonstrates that information in this category was most useful when connected to actions users could take.
Those actions included \action{Recognize potential choices}, \action{Request help with the system}, and \action{Modify algorithm operation}.

\paragraph{\infocategory{\textbf{AI Creators' Assumptions}}}
Participants sought information about AI Creators' expectations and assumptions about how users would work with or operate the AI system.
Critically, users looked to understand how AI Creators' assumptions aligned with critical aspects of their workflows. A correct assumption about the appropriate medical approach would make M2 ``go with the model, use it more because it’s helping me come up with a clinical practice that is the same line as what I would do.'' However, this Information Category was of limited use to some for taking actions.
M2 went on to note that even an incorrect assumption did not matter if the system ``still provides the right answer and my same line of thinking.''
On the other hand, T4 expressed that AI Creators' assumptions about users or their system did not matter so long as they received a favorable AI output. 
The most common action associated with this Information Category was \action{Decide/Justify usage of system given impacts}. 

\paragraph{\infocategory{\textbf{Data Sourcing and Use}}}
Participants sought information about the nature of the aggregate data sources used and the ways new data inputs are or aren't incorporated into the system. 
Participants often compared this information with their knowledge about data sources outside the system to anticipate downstream consequences of their decisions.
M5 applied knowledge from using other medical data to help them judge whether the system would operate well---``what was the sample size? Is this again, institution specific versus national? Were there any like, side effects or like, I guess, reasons behind choosing the alternatives, like cost, side effects, other pre-existing conditions contraindicated based on lab values.'' 
As a result, this Information Category led to actions about judging or deciding on their usage of the system. 
The top actions for this Information Category were Mental State Actions such as \action{Understand user’s or system’s impact on individual or wider society}, \action{Set expectations about an outcome}, \action{Recognize potential choices}, \action{Decide/Justify usage of system given impacts}, and \action{Understand broader contextual factors that affect the system}. 

\subsubsection{Information Categories related to User Environment} \label{Theme_UserEnv}

The four Information Categories under the theme of User Environment center on the broader environment shaping the AI system and users' interactions with it.

\paragraph{\infocategory{\textbf{Cross References}}}
The most frequently mentioned Information Category was \infocategory{Cross References}. Participants sought information from sources they trusted to add context to the system’s design, data inputs, data sources, outputs, and outcomes. For example, many participants relied heavily on a search bar in our interface that connected to research papers, teaching standards, and hospital- and school-specific information. 
Doctors expected to be able to turn to sources they trusted, such as PubMed, practice guidelines, or society-based recommendations (M2, M4, M5). Teachers also appreciated contextual resources, with one stating ``[this] chart is helpful when you're looking at a skill and right, here's where the state average is, and then so I can easily see where he's above, where he's the same, and where he's below'' (T1) and another appreciating ``the ability to look at different things that are there. I'm looking at different guidelines, state standards'' (T3).
These resources are not necessarily processed by the system or its algorithm, but users still wanted the information within the AI System to compare and judge the system's output with validated domain information. 
As with all Information Categories in this theme, the frequently associated actions connected the AI system to a broader context. The top Mental State actions for this Information Category were \action{Recognize Potential choices}, \action{Compare outputs from current model against other knowledge}, \action{Validate infor from current system}, and \action{Understand user's or system's impact on individual or wider society}. The top External actions were \action{Request more/different information}, \action{Consult another person about a decision}, and \action{Communicate about AI system information with another person}.

\paragraph{\infocategory{\textbf{Qualifications}}}
Participants sought information about the credibility of the AI Creators, the expert sources used to make the model, and the other system users. They used this information to assess the credibility and utility of the system. M5 asked who was involved in making the system to judge how much they should rely on its outputs, ``Who are the experts [creating the system]: Is it like the most well known endocrinologist in this area? Is it a pharmacist? Is it an interdisciplinary team?''
Other users' qualifications are also included in this category because they were important when system decisions included user input. 
M7 said, ``Seeing a 3rd year resident reject the [AI] recommended the medication might make you feel like you have a slightly different opinion than if it was rejected by a Pain Specialist [who has a lot more experience]...'' 
While the 3rd year resident and Pain Specialist made the same decision, the participant judged the qualifications differently, which lead to differing perceptions of the recommendations. 
This category was most frequently associated with Mental State Actions such as \action{Compare outputs from current model against other knowledge} and \action{Validate information from current system}. As with \infocategory{User Variables}, participants wanted information to connect with their domain contexts.

\paragraph{\infocategory{\textbf{Other Users' Experiences}}}
Participants sought information about other users' reasoning about, responses to, or actions based on the AI system. This Information Category includes whether other users accepted or rejected the AI system's recommendation and their reasons for doing so. 
For doctors, this information offer insights such as potential barriers to care that other doctors already encountered (M4, M5, M7, M8).
For teachers, other teachers' comments provide additional potential factors to consider about ``a student that looks the same on paper'' (T4).  
The information comparisons made and the ways they helped participants recognize relevant patterns across the data are reflected in many of the Mental State Actions taken: \action{Understand other peoples' choices/responses}, \action{Recognize potential choices}, \action{Compare the outcomes or outputs in my case to other outcomes or outputs}, \action{Understand user's or system's impact on an individual or wider society}, and \action{Recognize patterns across data}, among others.
These actions reflect the recurring importance of information for users to analyze consequences and downstream effects.

\paragraph{\infocategory{\textbf{Relationships with other Users}}}
Participants sought information about the other users making decisions and influencing the decision-making process. 
M7 needed information about other medical decision-makers, so that ``nothing that I'm doing kind of counteracts things what other people are doing''. 
T5 highlighted the importance of ensuring counselors consulted teachers prior to placing a student based on the AI's output. 
Participants took Mental State Actions to \action{Understand their decision-making responsibility} within the context of their relationships and influenced their ability to \action{Identify potential choices, options, and actions}. All participants also sought this information to know how to connect with \textit{other system users}, striving to take actions within and without the AI system to communicate with other people (\action{Share decision/rationale with another}; \action{Support another person's understanding}). 
As with other Information Categories, user actions focused on determining \textit{consequences} \textbf{and} \textit{context}. 

\subsubsection{Information Categories related to User Accountability} \label{Theme_UserAct}

The three Information Categories under the theme of User Accountability relate to users' options with and potential effects from use of the system. 

\paragraph{\infocategory{\textbf{Pathways for System Support}}} Participants sought information about their options and means to provide feedback or receive help. 
Importantly, this included information about what would happen after their feedback was given or help was requested. 
For example, teachers and doctors wanted to know the timeliness of system support. 
T3 explained that teachers need to quickly see updates to the system if an error was reported in order to avoid misplacing students in classes, 
``If I had to wait a semester, or any extended period of time, then I would, of course, I would lose faith in that because I need the recommendation now....So it would be super vital...that this information, that the prediction, I should say, was one based on the most current data, and that it was as accurate as we could get it.''  
M9 expressed similar sentiments around the timing for support in medical contexts, ``If it were medical, something medical, I would say 20 minutes [to get a response]. If you're asking [a] medical question [it] has to be quick. Financial stuff, you know, depends on how pressing it is. You'd like same day service... maybe an hour or so.'' 
This Information Category had the most External Actions associated with it, but it was not enough for system support pathways to be described. 
Participants once again wanted the information to include \textit{context} that would help them understand how system support had \textit{consequences} on the performance of their job and on stakeholders.

\paragraph{\infocategory{\textbf{Consequences for User}}}
This Information Category focuses on potential consequences for the user. T2 stated that this information could provide an ``objective'' perspective to balance out personal worries about hypothetical consequences. 
M3 felt, however, that information about consequences drawn from non-validated means such as other users 
could perpetuate anecdotal data over evidence-based information.
Mental State Actions reflected the \textit{contextual} as well as consequential implications of this kind of information, such as allowing participants to \action{Understand user’s or system’s impact on individual or wider society}, \action{Set expectations about an outcome}, and \action{Counter harm (malicious or unintentional)}.

\paragraph{\textbf{\infocategory{Consequences with Other Stakeholders}}}
Participants sought information about impacts on users' interactions with other stakeholders, such as patients, caregivers, students, and parents.
Teachers used this information to add ``the human component'' (T1) of educational decision making while using the AI system. 
Doctors used the information to better understand their patients' healthcare needs and priorities (M3) and communicate with caregivers such as patients' medical power of attorneys, spouses, or children (M7). For example, M8 expressed they wanted to see information about costs for medications alongside AI outputs to judge how affordable and sustainable a treatments would be for patients. 
This category was second only to \infocategory{Pathways for System Support} for total External Actions used. Further, these actions most often involved interacting with another person, demonstrating the importance and relevance of system-provided information to users' relationships \textit{outside the system}.

\section{Applying the Catalog} \label{Extended_Example}
Our work advances a core goal of AI explanations---helping users do something in response to an AI system---by synthesizing a catalog of information for evaluating AI systems. The catalog supports the measurement and evaluation of AI explanations in terms of users' actions through user-centered terminology, information categories, and action types. 
In this section, we use a running example to illustrate how to apply the catalog to evaluate and design for AI system actionability \textit{with users}. 
The example is based on prior work about a deployed AI system to help welfare workers identify cases where children may be at-risk of violence   ~\cite{Stapleton2022_Extended,ChengEtAl2022}.
The system included a black box model and an XAI system to help users understand the reasoning behind AI decisions.  We chose this example as a third domain to demonstrate the use of our catalog beyond the two study domains.

\subsection{Actionable Feedback on (X)AI systems with User-Centered Terminology}
AI Creators conduct evaluations to receive user feedback that can inform system improvements. 
The catalog's user-centered terminology ensures clear communication with users and study participants. 
Current predominant approaches for evaluating ``black box'' models often assume users understand and will distinguish between the AI system, AI model or agent, and the explanation of the system \cite{Yao2021,LimEtAl2019}. 
However, as discussed before, our findings suggest that users do not distinguish between these terms and that making this assumption may lead to unclear feedback.  

If AI Creators project their understanding of the system onto users, they can misinterpret users’ feedback on the actionability of a system. For our example AI system, a welfare worker may say that an AI system does not fit their process of identifying at-risk children, intending to indicate they cannot use the system to change a decision about which child to screen (actionability). Without understanding that users may likely be referring to the entire AI system---AI, XAI, interface, and any other information---AI Creators may instead interpret the user’s statement to refer to the AI model, the variables or performance of the underlying algorithm performing the prediction. In this way, using  and relying on language in a way that reflects users' understanding can clarify what aspects of the AI model, explanation, or application that must change.

\subsection{Evaluating Actionability of AI Systems with User-Centered Action Types}

Our catalog provides actions that AI Creators can put at the center of their evaluations. We identified more than 60 User-Centered Action Types across 3 dimensions of actionability---AI Interactions, External Actions, and Mental State Actions. 
While the list of actions is not exhaustive, it is not infinite either, lending itself to an iterative process to probe, explore, and address users' needs. 
AI Creators can use these action types to express which user actionability needs they address. 
In the AI system for welfare workers, 
an AI Creator may design the system for welfare workers to \action{Change a decision about an action} with a system-provided cumulative risk score indicating which cases should be investigated.
AI Creators can then assess whether users take that action. 

Further, our work captures three actionability dimensions: Mental State Actions, AI Interactions, and External Actions. 
As highlighted in other studies \cite{JornoAndGynther2018,FanniEtAl2022}, Mental State Actions impact how people coordinate and act in the physical world. Considering Mental State Actions alongside AI Interactions and External Actions can help AI Creators better specify how explanations change what people know, allowing them to act in the response to an AI system. 
For example, AI Creators working on the AI system for welfare workers can reference the catalog to explore and identify a broader range of users' Mental State Actions such as \action{Recognize patterns in the data}, \action{Understand the variables and data used to make the system}, \action{Compare the outputs in one case  to other outcomes or outputs}, and \action{Judge model performance}. Further, they can connect Mental State Actions to AI Interactions, such as \action{Request more information}, and External Actions, such as \action{Consult another person about a decision}. 

The breadth and granularity of our catalog's actions can guide AI Creators as they iterate on and update quantitative and qualitative evaluations. 
In the welfare example, AI Creators may assume welfare workers \action{Change a decision about an action} but find that few welfare workers report that as their intended action.
Welfare workers, in interviews or open ended-responses, might instead describe their need for External Actions such as \action{Consult another person about a decision} and \action{Share a decision/rational with other people} because of the broader process of investigating a child's case. 
Based on these insights, AI Creators can then iterate on the designs of their systems. 
This helps AI Creators understand how the AI system is situated in users' broader context while uncovering areas where information in the AI system should be adapted. Because this analysis would center on users' actions, the findings could then once again be re-framed as an evaluation for the AI system, more deeply integrating users' perspectives.

\subsection{Layering Information for Actionability with User-Centered Information Categories}

The catalog's Information Categories structure the range of AI System information that users may rely on to take action. The four Information Category themes reflect users' desires to draw together information from across the AI model and their environment to make decisions.

In the welfare worker example, AI Creators can specify which Information Categories they are incorporating and then analyze welfare workers' actions to evaluate if and they use this information. By listing the most frequent actions our participants associated with each Information Category, our catalog provides a starting place to adapt explanations based on feedback in evaluations. This may look like 
embedding system decisions alongside information detailing \infocategory{Other Users' Experiences} and \infocategory{Qualifications}, such as information about decisions made by co-workers and their prior experience making similar choices.  
Providing information about \infocategory{Consequences with Other Stakeholders} and \infocategory{Alternative Outcomes with the Model}, such as prior health issues that the child has had and the outcomes of children with similar issues, may help welfare workers understand the potential impacts of the AI decision on the children and families who are investigated. 
AI Creators could then examine the actions that AI system's enable when they offer these Information Categories to ensure welfare workers' can appropriately leverage the AI system. 

\section{Discussion} \label{Discussion}
Measuring, evaluating, and improving the actionability of explanations is critical for creating impactful AI and XAI systems that help---not hinder---human decision makers.
People in high-stakes decision-making contexts, such as hospitals, schools, and financial institutions integrate AI systems into their daily decision making. Ideally, AI explanations enable people to act in response to an AI system by changing what they know~\cite{Liu2021,Silva2019,Coeckelbergh2020}. 
Previous studies showed, however, that AI explanations are not always actionable,
and consensus did not previously exist about the information that effectively enables user action.

Our catalog presents information categories for explanations and three dimensions of actionability taken with those explanations. 
The catalog's design includes:
\begin{itemize}
    \item \textbf{User-Centered Terminology}, which enables clearer communication with users;
    \item \textbf{User-Centered Action Types}, which offer a breadth of 60 actions separated into three kinds of actionability---13 unique AI Interactions, 17 unique External Actions, and 30 unique Mental State Actions---that users may need to respond to an AI system; and 
    \item \textbf{User-Centered Information Categories}, which describe 12 different kinds of information that can be mapped to the actions they enable.
\end{itemize}
Our catalog serves as a practical tool for investigating and designing explanations that meet users' complex and dynamic needs, and it accelerates progress towards XAI's core assumption---that explanations enable action---in multiple ways. 

\subsection{Articulating the Connection Between Information and Action}
AI Creators can use the catalog to articulate ``\textit{what users understand explanations tell them} and how \textit{they act} upon these understandings'' \cite{ChenEtAl2022}. 
As demonstrated in our study, users' knowledge and assumptions about AI Systems differ from those of AI Creators, which is reflected in the language they use around actions and explanations. 
For example, users' do not distinguish between the AI system and an explanation system. 
This is currently at odds with predominant approaches for evaluating ``black box'' models, where evaluations assume users understand and distinguish between the AI system, AI model or agent, and the explanation of the system \cite{Yao2021,LimEtAl2019}. 
The catalog translates between AI Creators' understanding of their systems and users' understanding of these systems.
The catalog offers a wide design space for an iterative, evaluative process to probe, explore, and address users' needs. 
The result is productive evaluations with clearer feedback that lead to more actionable AI Systems for users.

Additionally, we describe how participants followed a ``nested'' information seeking pattern that compares model recommendations to other trusted sources. For example, doctors expected \infocategory{Cross References} from domain-trusted sources, such as PubMed, practice guidelines, and society-based recommendations (M2, M4, M5), to compare against the system-provided information. Participants relied on these comparison to understand how AI outcomes could be incorporated into their decision-making. This confirms and extends prior work on AI explanations demonstrating the need for explanations to connect to doctors' work flows \cite{CortiEtAl2024}, and highlights the need to integrate multiple kinds of information in order for AI systems to be actionable.

\subsection{Addressing Mental State Actions---an Under-Explored Dimension of Actionability}
For the first time that we are aware, our work provides \textbf{empirical evidence} of users themselves articulating the existence of {\em Mental State Actions} as critical and impactful for users. Participants described Mental State Actions---such as, brainstorming different options, understanding what the model is looking for, persevering or feeling motivated, getting angry, and setting different goals, among others---as \textit{valuable actions} that impacted how they took physical actions in their environment, including how they would talk to patients or other teachers and the kinds of information they'd look for in the interface. Mental State Actions were also the most frequently taken, as participants requested Mental State Actions 2190 times, almost 10 times greater than the 200 AI Interactions and 270 External Actions they would take. Our finding supports Jørnø \& Gynther  \cite{JornoAndGynther2018} and Fanni et al. \cite{FanniEtAl2022} who theorize that an actor's ``action capabilities'' encompass \textit{mental actions} or choices, which can impact physical actions or interactions. 
Other XAI researchers have argued for mental state actions such as thinking, planning, reflecting \cite{NeffAndNagy2018,SidorovaAndRafiee2019,LangleyEtAl2017,Swanepoel2021,HarrellAndZhu2009,VannesteAndPuranam2023,Dattathrani2023,Longin2020,List2021,Deschenes2020,GibbsEtAl2021,AdenugaAndDodge2023}, and making choices \cite{CemalogluEtAl2019,Murray2021,Schonau2022,KhosraviEtAl2022,FanniEtAl2022,Swanepoel2021,HarrellAndZhu2009,List2021,Deschenes2020,GibbsEtAl2021,Langley2019} because they help people coordinate and direct actions in the physical world. Our work presents empirical evidence for this a previously under-explored dimension of actionability in evaluations.

Our catalog also records the variety of users' Mental State Actions at a level of granularity that makes these difficult-to-observe actions more easily specified and evaluated.
AI Creators can leverage the granularity of the actions we list to identify and evaluate the connection between users' Mental State actions, AI Interactions, and External actions. 
For example, on a survey, AI Creators can ask users to select or rank actions they felt they could take because of the explanation from a mixed list of Mental State Actions, AI Interactions, and External actions. 
When analyzing transcribed interview data, AI Creators could also qualitatively code which Mental State actions are mentioned by participants and how they connect to other actions they would like to take with the explanations. Because our catalog surfaces these actions, they can be more easily measured and integrated into the design of AI systems.

\subsection{Comparing Findings Across the Community}
The catalog benefits the XAI and AI communities as a whole. 
Due to previous lack of consensus on terminology, it was difficult to transfer findings in XAI literature. 
For example, there were differences in what is recognized as an action \cite{MothilalEtAl2020,SinghEtAl2021,GhoshEtAl2018,JornoAndGynther2018}, what information constitutes or can be offered in an explanation \cite{ArrietaEtAl2019,ICOAndTuring2022,LimEtAl2019,Lipton2018,EdwardsAndVeale2017}, and what was part of an interface, XAI system, or AI system \cite{YangEtAl2019,CortiEtAl2024,ChariEtAl2020,Lipton2018,RobertsEtAl2012}.  
The unified, user-centered terminology, information, and actions in our catalog help AI Creators understand, compare, and integrate findings across studies in the field of XAI.

\section{Limitations and Future Work} \label{Limitations}
Our work is a formative step towards helping AI Creators evaluate the actionability of their explanations \textit{with users}. While the scenario-based design activity provided a user-centered approach, we acknowledge the method's limitations, including its dependency on the scenario created, the strength of the connection between participants and the scenario, and the variability of participants' experiences. We address some limitations through recruiting of participants with appropriate experience and through careful, iterative design of the study visuals. 
However, insights should be viewed as formative. We highlight the commonalities across the two professions in our study, but we acknowledge that differences existed and likely would extend in other decision-making domains. We aim to further investigate similarities and differences of other domains in future work. Finally, our work focused largely on information and explanations provided in written text. There are contexts (e.g. radiology) where visual explanations such as images are important. We acknowledge the need to expand the design space to explanations in multiple modalities.

\section{Conclusions} \label{Conclusion}

Evaluations are a critical part of how AI Creators ensure the effectiveness, safety, and proper functioning of AI systems. 
Our understanding of the information and actions that users desire and associated algorithmic developments was limited by the lack of a systemic catalog of information and actions for AI Creators to articulate and compare explanation approaches. 
Through our study, we interviewed 14 participants in two domains and contribute a catalog of information and actions to guide evaluations of AI systems. 
Our catalog's User-Centered Terminology, Information Categories, and Action Types serve as a rich source for AI Creators to evaluate and design how their AI systems enable user actions. We provide an example of applying the catalog to evaluate AI Systems with users. 
The catalog's structure reflects users' priorities around actionability, helping AI Creators anticipate and design for the actions enabled with system explanations to truly center users and actionability.

\begin{credits}
\subsubsection{\ackname} 

The author gratefully acknowledges  the support of the National Science Foundation GRFP under Grant No. DGE-2039655. Any opinion, findings, and conclusions or recommendations expressed in this material are those of the authors(s) and do not necessarily reflect the views of the National Science Foundation. 

\subsubsection{\discintname}
The authors have no competing interests to declare that are
relevant to the content of this article.
\end{credits}
%
%
%
\bibliographystyle{splncs04}
\bibliography{references}

\newpage


\appendix \label{Appendix}

\section{Additional User-Based Terminology} \label{Appendix_Terminology}
As discussed in Section \ref{analysis}, we developed user-based terminology and definitions, which we used in coding our data and throughout the rest of the paper. Here we provide a full description of the definitions we derived. 

Participants use \textit{AI system} (or \textit{system} for short) to refer to the AI algorithm \textbf{and its corresponding user interface}, including all explanations and elements that are not directly tied to the AI algorithm, such as a ``contact us`` button. They use the term \textit{model} to specifically denote the AI algorithm. The \textit{AI Creators} are all people who were involved in making the AI system, both its model and interface. This includes software developers, the design team, and any other consultants or people who helped created and test the system, not just the AI model creators. These user-derived terms clearly diverge from the AI community's tendency to define them solely in the context of the AI-specific portions of a larger application.

The \textit{user} is the person directly interacting with the AI system who has to make \textit{decisions}, or choices, when using the AI system. The output is distinct from the \textit{outcome}, which is the resulting situation or consequences from a human decision or AI output. Finally \textit{stakeholders} are all persons impacted by the user's decision making or AI system's outputs. 

For example, the AI's recommendation of a 30 mg Penicillin dose is the \textit{output}; the doctor who directly uses the AI system is the \textit{user}, and their choice to prescribe a 50 mg Penicillin dose is the \textit{decision}; the patient and their caregivers are both \textit{stakeholders}, and the patient's reduced pain is the outcome.

Other content displayed in the system but not generated by the AI, such as contact information, is referred to broadly as \textit{information}. 
With regards to composition of the AI system, \textit{Data inputs} are the raw data used to make the model's AI algorithms. The \textit{Data source} is where the data came from, and \textit{variables} are the parts of the data inputs used to make the AI model. If the data inputs could be likened to a yard of fabric, then the data source is analogous to Joann's, and the variables are 3 squares cut and used from the fabric to make a pencil holder. As AI creators make decisions around their data sources and inputs, they make \textit{assumptions}, accepting as true some things about how the system will be used or should operate within a context. These assumptions ultimately constitute the AI system's \textit{implementation}, or choices about how the model works that impact what the overall AI system can do and how it can be updated or changed.

\newpage
\section{Action Codes} \label{Appendix_ActionCodes}

\begin{table}[h!]
\centering
\begin{tabular}{||p{4cm} p{9.5 cm}||} 
 \hline
 \textbf{Mental State Action} & \textbf{Description} \\  
 \hline
Change a decision about an action & User chooses a different action and/or behavior than the one they were going to take/exhibit, including doing something when they were going to do nothing and vice versa \\ 
\hline
Choose one option over another & Decides among multiple options, including prioritization \\
\hline
Emote or feel things about the system & Explicitly expresses emotions or feelings about the system, its creation, and/or other users (e.g. desire, interest, enjoyment, or dislike, trust, confidence) \\
\hline
Compare the outcomes or outputs in one case to other outcomes or outputs & Compares and contrasts one set of variables, outputs, outcomes, or other system-provided information with others within the system (can be other users or stakeholders) \\
\hline
Recognize patterns across data & Analyzes and synthesizes a set of system-provided outcomes or data points \\
\hline
Counter harm (malicious or unintentional) & Considers factors, actions, or other ways to prevent or respond to a system’s harm or negative effects (whether intentional or not) including bias \\
\hline
Decide/Justify usage of system given impacts & Chooses whether to use the system based on an assessment of its effects \\
\hline
Assess a person’s credibility or qualifications & Considers, maintains, and/or decides on the credibility or qualifications of oneself or others \\
\hline
Understand variables and/or data used to make the system & Reviews and processes data and/or variables used by the AI model or the interface \\
\hline
Understand implementation and design as it relates to bias & Reviews and processes system design choices or considerations, such as (un)used data or variables, that affect the user’s perception of bias \\
\hline
Judge model performance & Reviews the model’s metrics and assesses its performance \\
\hline 
Know possible outcomes & Understands or identifies at least two potential system outcomes or outputs \\
\hline
Know who, how, or when to contact & Understands the people or means for getting help with the system or their decision, such as AI Creators, tech support, other users, or consultants \\
\hline
Recognize potential choices & Identifies options and ways to change those options, including actions that could be taken and information that could be considered \\
\hline
Recognize outputs/products made by AI & Understands that a system artifact was created by AI (e.g. AI-composed medical note, AI-created code snippet) \\
\hline
\end{tabular}
\caption{Action Codes and Descriptions for Mental State Actions; continued on next page}
\end{table}

\begin{table}[h!]
\centering
\begin{tabular}{||p{4cm} p{9.5 cm}||} 
 \hline
 \textbf{Mental State Action} & \textbf{Description} \\  
 \hline
Set expectations about an outcome & Identifies or develops a belief based on system information that outcome X will occur when Y conditions are met (does not include setting expectations about system support) \\
\hline
Uncover other areas to explore & Identifies new information beyond the model inputs and outputs for review, including additional data inputs and codified external information such as PubMed or the College Board \\
\hline
Understand broader contextual factors that affect the system & Identifies and assesses knowledge from outside the system that contextualizes the system’s data, outputs, or outcomes, such as sensor error rates, limits of standardized testing, and causes of systemic injustices \\
\hline
Understand user’s or system’s impact on individual or wider society & Identifies first-order effects that the user’s decisions and/or system-related outputs and outcomes will have on the user themselves, stakeholders, or society at large, such as a medication that will be a ‘barrier to care’ for certain patients (does not include effects of systemic factors such as bias)\\
\hline
Understand model algorithm & Identifies and assesses the AI Creators’ algorithm design, such as resources informing the algorithm (e.g. the drug dosing guidelines) and model construction (e.g. the weights of model parameters) \\
\hline
Understand other peoples’ choices/responses & Considers and assesses system-provided information about another person, such as their decision or rationale \\
\hline
Understand system assumptions & Identifies AI Creators’ assumptions about the system’s context, users, and use based on explicit system components such as ‘About’ text and/or implicit system components such as information excluded and roles represented \\
\hline
Understand the recommendation & Literally knows the meaning of the system’s output (e.g. understanding the recommendation to raise your credit score) \\
\hline
Understand the data that influence the output & Connects the variables, data inputs, and other system information and reasoning to the system output \\
\hline
Understand what is not included in the model/system and/or why & Knows system excluded elements, including variables not considered and reason for exclusion \\
\hline
Understand/verify credentials of those referenced/involved & Reviews and/or assesses the qualities of AI creators, other users, system inputs, and sources cited \\
\hline
Validate information from current system & Assesses system information (e.g. other users’ inputs, ‘factual’ information, and outputs) against other knowledge  \\
\hline
Understand decision-making responsibilities & Assesses a person’s sphere of control, permissions, and/or rights regarding actions and decisions (includes user or others stakeholders) \\
\hline
\end{tabular}
\caption*{Action Codes and Descriptions for Mental State Actions; continued from previous page}

\end{table}

\newpage

\begin{table}[h!]
\centering
\begin{tabular}{||p{4cm} p{9.5 cm}||} 
 \hline
 \textbf{AI Interaction} & \textbf{Description} \\  
 \hline
 Agree/Consent to AI rationale or decision & Presses a button or otherwise directly interacts with the system to accept a system-provided outcome (can be combined with ‘Edit an AI response’) \\
 \hline 
 Communicate about AI system information with another person & Uses the system to send system-provided information (e.g. an output, text, artifacts) to another person \\
 \hline 
 Consult another person about a decision & Uses the system to solicit input from another person about an issue or decision \\
 \hline
Correct information/data & Uses the system to correct an error in system-provided information \\
 \hline 
 Edit an AI response & Uses the system to tailor system-provided information (does not include corrections) \\
 \hline 
 Export/print information from an AI system & Uses the system to create products (e.g. PDFs, graphics, text) that can be used outside the system, such as printing a hard-copy or showing in a presentation to others \\
 \hline
 Modify algorithm operation & Uses the system to change the way the AI model works; includes modifying the data inputs (coded with ‘Correct data’ or ‘Recollect/input new data’) and model design parameters such as variable weights \\
 \hline 
 Input new data & Uses the system to input new data inputs into the system (does not include modifying existing information) \\
 \hline 
 Request system support & Uses the system to request assistance with the system (e.g. reporting missing information, bias, errors; asking for an audit)\\
 \hline
 Request more/different information & Uses the system to ask for more information beyond what is already provided (e.g. additional comparisons, presentations of data, etc.) \\
 \hline
Request a different presentation mode & Uses the system to ask for a different view of existing system-provided information (e.g. bar chart instead of text) \\
\hline 
Share a decision/rationale with other people & Uses the system to inform other individuals or groups about a decision or rationale (e.g. direct message, forum posts, etc) \\
\hline
Explore a raw dataset & Uses the system to examine the system’s raw data inputs \\
\hline 
 \end{tabular}
 \caption{Action Codes and Descriptions for AI Interactions}
 \end{table}

 \newpage

\begin{table}[h!]
\centering
\begin{tabular}{||p{4cm} p{9.5 cm}||} 
 \hline
 \textbf{External Action} & \textbf{Description} \\  
 \hline
 Change state & Alters their own mental or physical state based on the system (yes, it’s a broad category) \\
 \hline 
 Communicate decision-making responsibility & Communicates outside the system with another person about decision-making responsibilities \\
 \hline 
 Communicate system information to/with another person & Communicates outside the system with another person about system-provided information \\
 \hline 
 Conduct academic/clinical research & Formally investigates an issue through academic or clinical activities outside the system (i.e. a study or other project; not reading a single paper to be more informed) \\
 \hline 
 Consult another person about a decision & Solicits outside the system for input (e.g. opinions, advice, approval) \\
 \hline 
 Counter harm (malicious or unintentional) & Acts outside the system to address a system’s harm or negative effects (whether intentional or not) including bias \\
 \hline 
 Maintain credibility/qualifications of a person & Acts outside the system to preserve or otherwise consider the credibility or qualifications of oneself or others \\
 \hline 
 Influence legal regulation & Acts outside the system to change or otherwise influence regulations (e.g. protest, contacting representatives) \\
 \hline 
 Prevent/Discourage a behavior & Acts outside the system to decrease the likelihood of a behavior by oneself or others \\
 \hline 
 Rally others to help make an outcome happen & Acts outside the system to encourage others towards a desired outcome \\
 \hline 
 (Re)Collect new data & Acts outside the system to acquire data inputs not included in the system \\
\hline 
Request system support & Requests assistance outside the system on a topic related to the system (e.g. reporting missing information, bias, errors; asking for an audit) \\ 
\hline 
Request more/different information & Seeks more information outside the system beyond what is already provided (e.g. web searches, research databases, talking to another person) \\
\hline 
Share a decision/rationale with another person & Acts outside the system to inform other individuals or groups of a decision or rationale (e.g. direct message, forum posts, etc.) \\
\hline
Stop using the system & Decides to not use the system (can include switching to a different system) \\
\hline
Support another person’s understanding of the system output & Shares information outside the system to help another person’s understanding \\
\hline
Teach & Acts outside the system to teach or educate (must be explicitly stated) \\
\hline 
 \end{tabular}
 \caption{Action Codes and Descriptions for External Actions; continued from previous page}
 \end{table}

\clearpage

\section{The Full framework} \label{Appendix_FullFramework}
Here we provide the full set of actions associated with all Information Categories. For each Information Category we distinguish between the high, medium, and low frequency actions.  To determine which actions were in each category, we calculated the total number of each actions coded for each Information Category. High frequency actions (underlined) were in the top quartile of the totals. Medium frequency actions (italicized) were in the second quartile. Low frequency actions were in the bottom quartile and were coded at least once. For readability, we present the framework by theme (Model Exposure, Model Creation, User Environment, User Accountability). We distinguish between Mental State, AI Interactions, and External Actions categories.

\begin{figure}[t]
 \includegraphics[width=\textwidth]{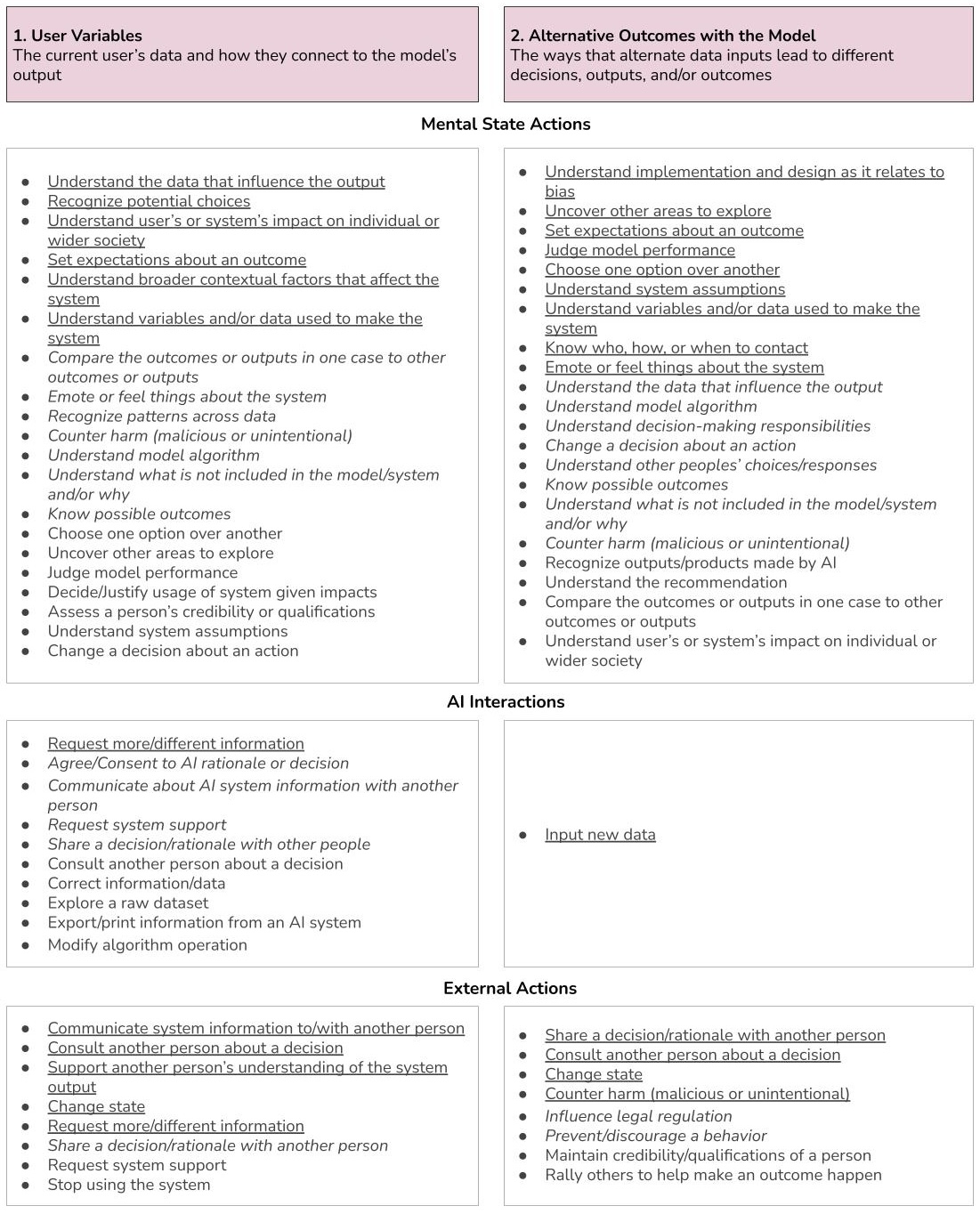}
   \caption{Information Categories under Model Exposure and \textit{all} actions associated with them.}
   \label{fig:Framework_ModelExposure-FULL} 
\end{figure}

\begin{figure}[t]
 \includegraphics[width=1.2\textwidth]{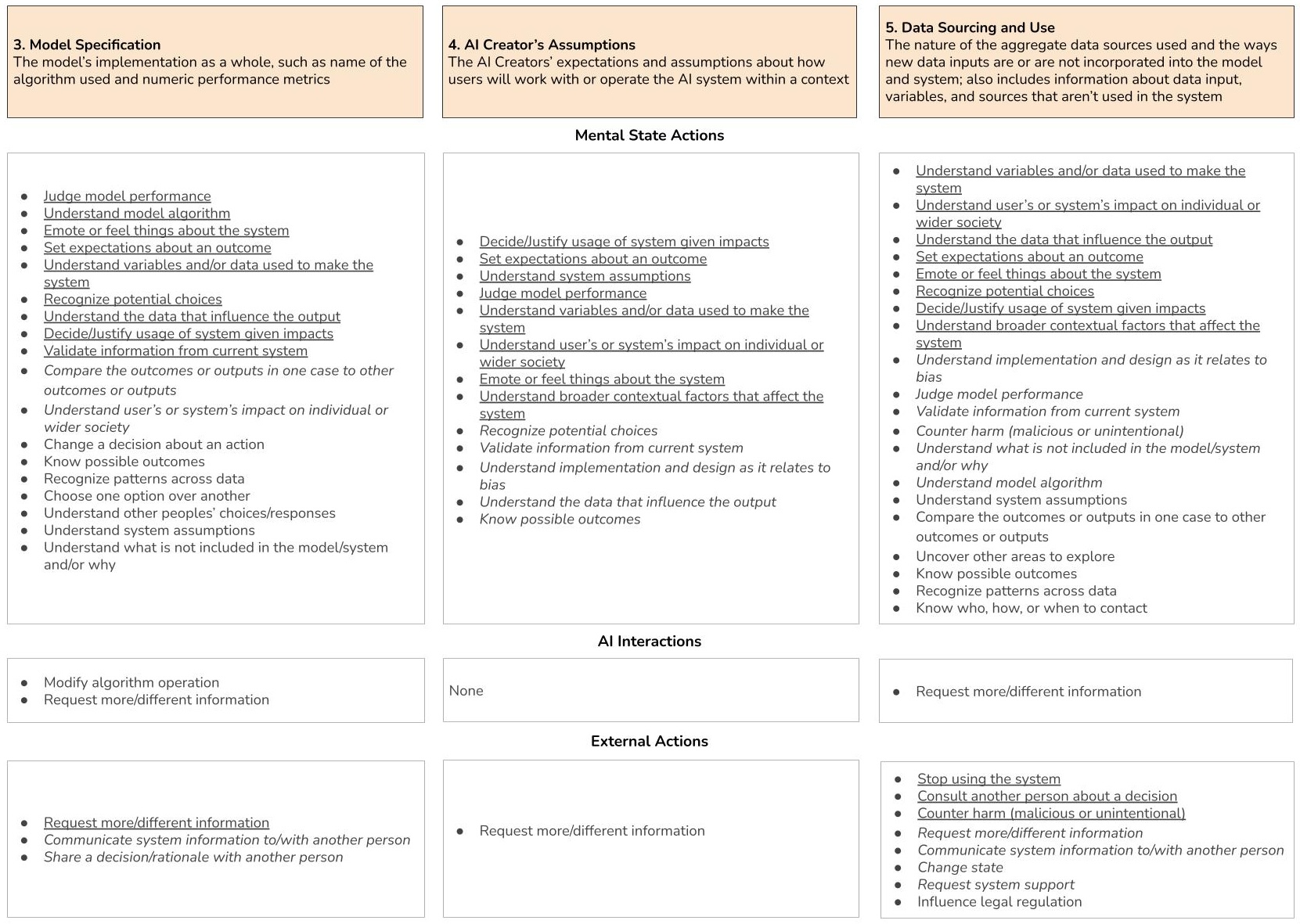}
   \caption{Information Categories under Model Creation and \textit{all} actions associated with them.} 
   \label{fig:Framework_ModelCreation-FULL} 
\end{figure}

\begin{figure}[t]
 \includegraphics[width=1.2\textwidth]{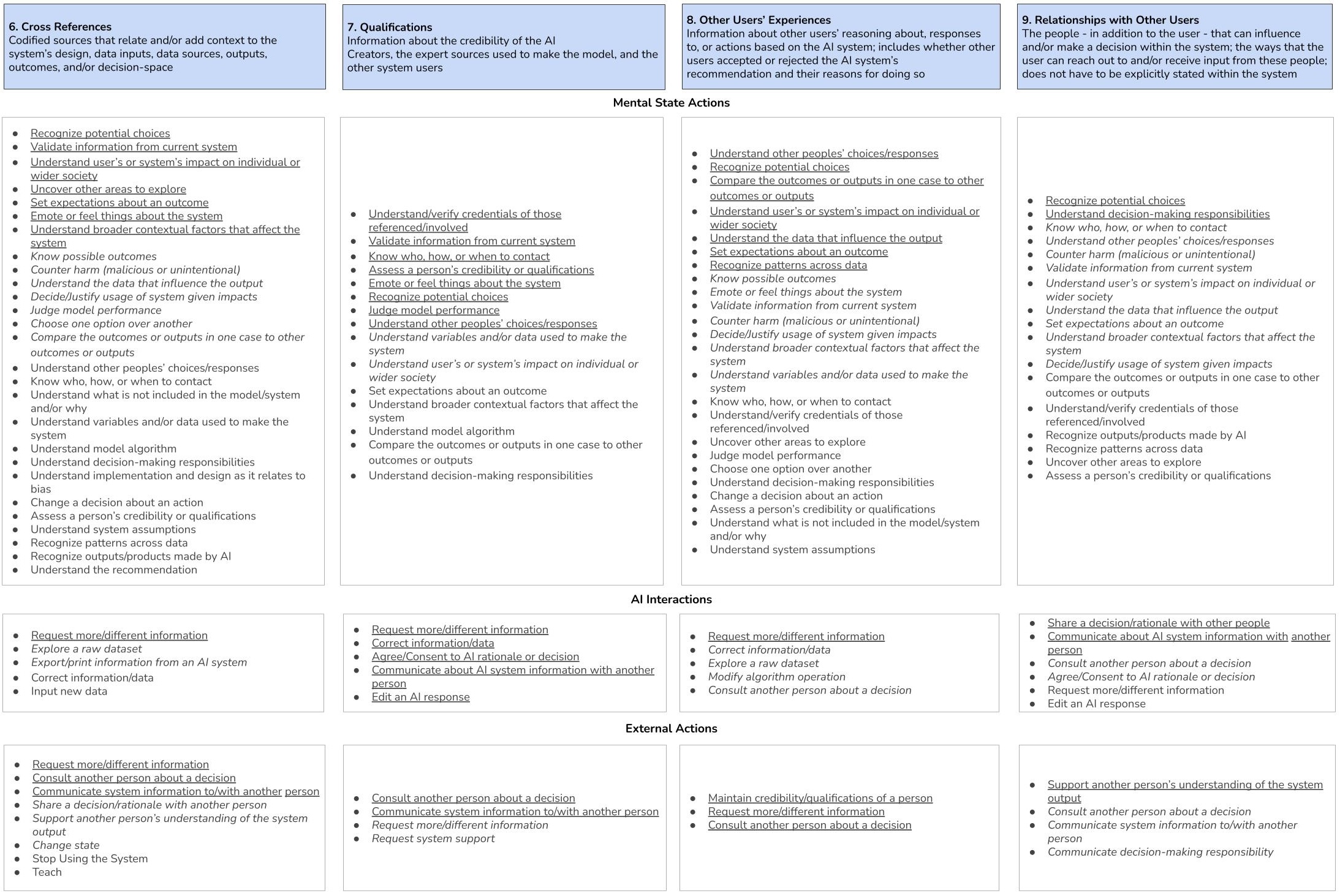}
   \caption{Information Categories under User Environment and \textit{all} actions associated with them.} 
   \label{fig:Framework_UserEnvironment-FULL} 
\end{figure}

\begin{figure}[t]
 \includegraphics[width=1.2\textwidth]{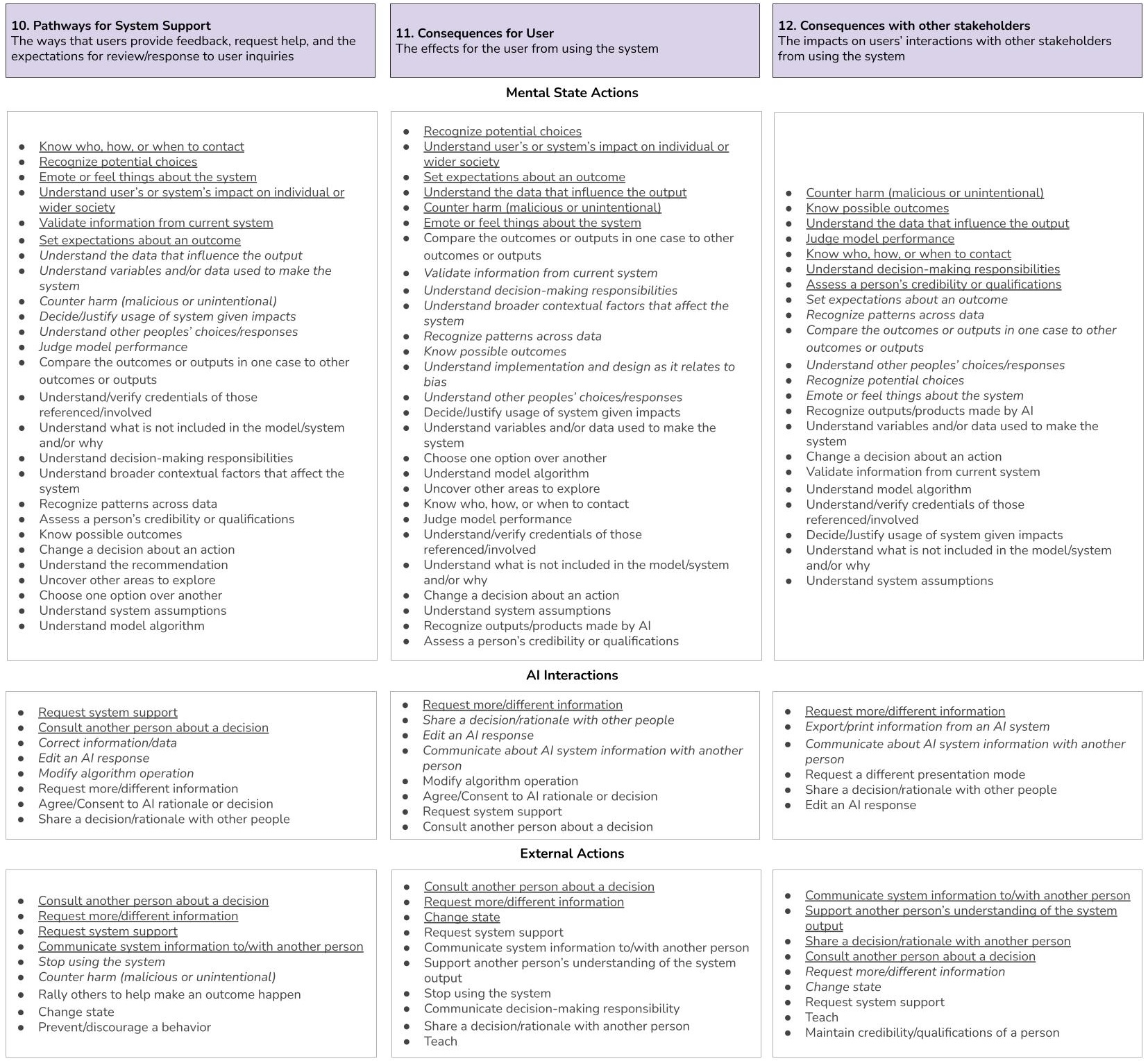} 
   \caption{Information Categories under User Accountability and \textit{all} actions associated with them.}
   \label{fig:Framework_UserAccountability-FULL} 
\end{figure}

\end{document}